\documentclass{article}
\usepackage{spconf,amsmath,graphicx}

\usepackage[labelformat=simple]{subcaption}

\usepackage[font=small,labelfont=bf]{caption}
\usepackage{multirow}
\usepackage{adjustbox}
\usepackage{amssymb}
\usepackage{bbm}
\usepackage{booktabs}
\usepackage{float}
\usepackage{amsthm}
\usepackage{color, colortbl}

\usepackage{pifont}

\definecolor{Gray}{gray}{0.9}



\allowdisplaybreaks

\title{Continuous Streaming Multi-Talker ASR with Dual-path Transducers}
\name{Desh Raj\sthanks{Work done during internship at Microsoft.}$^1$, Liang Lu$^2$, Zhuo Chen$^2$, Yashesh Gaur$^2$, Jinyu Li$^2$}
\address{$^1$Center for Language and Speech Processing, Johns Hopkins University, USA, $^2$Microsoft Corp., USA}
\begin{document}
%
\maketitle
\begin{abstract}
Streaming recognition of multi-talker conversations has so far been evaluated only for 2-speaker \textit{single-turn} sessions. In this paper, we investigate it for \textit{multi-turn} meetings containing multiple speakers using the Streaming Unmixing and Recognition Transducer (SURT) model, and show that naively extending the single-turn model to this harder setting incurs a performance penalty. As a solution, we propose the dual-path (DP) modeling strategy first used for time-domain speech separation. We experiment with LSTM and Transformer based DP models, and show that they improve word error rate (WER) performance while yielding faster convergence. We also explore training strategies such as chunk width randomization and curriculum learning for these models, and demonstrate their importance through ablation studies. Finally, we evaluate our models on the LibriCSS meeting data, where they perform competitively with offline separation-based methods. 
\end{abstract}
\vspace{-0.5em}
\section{Introduction}
\label{sec:intro}
\vspace{-0.8em}

With advancements in clean, single-speaker transcription~\cite{Amodei2016DS2,Xiong2017TowardHP}, researchers are focusing on harder speech recognition settings with interference in the form of noise, reverberation, or overlapping speakers~\cite{Barker2015TheT, Kinoshita2013TheRC, Watanabe2020CHiME6CT}. Of these, the latter is particularly irksome, while also being prevalent in meetings and natural conversations~\cite{Carletta2005TheAM, Shriberg2001ObservationsOO, Yoshioka2019MeetingTU}. The conventional approach to handle overlapping speech is through a cascade of separation and recognition modules, which may be sub-optimal since the modules are independently optimized. It also requires considerable engineering efforts to maintain the pipeline. 

An alternative solution to this problem is through the application of end-to-end models that are trained with text-based supervision~\cite{Graves2006ConnectionistTC, Graves2012SequenceTW,Lu2016OnTT, Chorowski2015AttentionBasedMF, Chiu2018StateoftheArtSR, he2019streaming, Li2019RNNT, li2021recent}. Several model paradigms have been explored, including attention-based encoder-decoders~\cite{Settle2018EndtoEndMS, Chang2019EndtoendMM, Kanda2020SerializedOT}, and recurrent neural network transducers (RNN-T)~\cite{Tripathi2020EndToEndMO, Sklyar2021StreamingMA}, with most of the models functioning in an offline mode. For the streaming case, Lu et al. recently proposed Streaming Unmixing and Recognition Transducer (SURT)~\cite{Lu2021StreamingEM} which simultaneously transcribes overlapping speech into two channels through ``unmixing'' and ``transcription'' modules that are jointly optimized using an RNN-T based heuristic error assignment training (HEAT) loss. We will review SURT in \S~\ref{sec:surt}.

SURT was originally evaluated on 2-speaker single-turn sessions and showed promising results. However, real-life meetings often contain multiple speakers and several turns of conversation, and it is not immediately clear whether the ``vanilla'' SURT model would transfer well to these harder settings. In this paper, we show that the train-test mismatch does, in fact, incur performance costs that cannot be easily recovered using multi-turn training. We then investigate dual-path (DP) LSTM and Transformer models, which have been successfully applied in speech separation~\cite{Luo2020DualPathRE, Chen2020DualPathTN}, and show that they are better at modeling the longer sequences that occur in multi-turn meetings. Our contributions include training strategies for DP models, such as curriculum learning and chunk width randomization, and a theoretical analysis of DP Transformers. We evaluate our models on several tiers of simulated multi-talker mixtures and on the LibriCSS meeting corpus~\cite{Chen2020ContinuousSS}, where it performs competitively with offline separation-based approaches.

\begin{figure}[t]
\centering
\includegraphics[width=\linewidth]{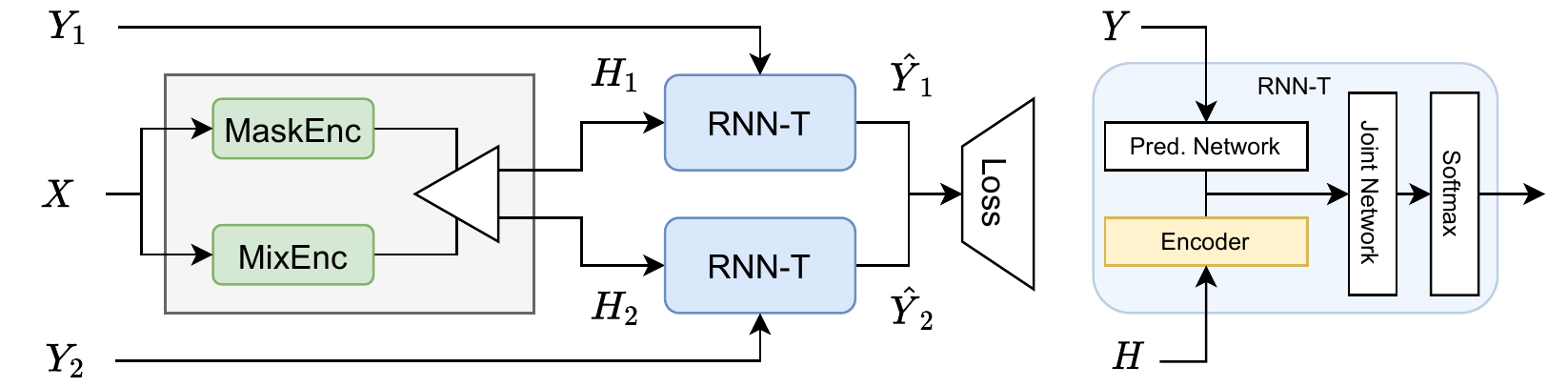}
\caption{An overview of the RNN-T based SURT model for the 2-speaker overlapping case.}
\label{fig:surt}
\vspace{-1em}
\end{figure}

\vspace{-0.5em}
\section{SURT: Review}
\label{sec:surt}
\vspace{-0.8em}

Fig.~\ref{fig:surt} shows an overview of the RNN-T based SURT model. Given input features $X$ obtained from an overlapping speech session, a mask-based unmixing module extracts speaker-dependent representations $H_1$ and $H_2$ as
\begin{align}
& H_1 = M \ast \bar{X}, \quad H_2 = (\mathbbm{1} - M) \ast \bar{X}, \\
& M = \sigma(\texttt{MaskEnc}(X)) ~ \text{and} ~ \bar{X} = \texttt{MixEnc}(X). \nonumber
\end{align}
Here, $\sigma$ denotes the Sigmoid function, $\ast$ is the Hadamard product, and $\mathbbm{1}$ has the same shape as $M$. Both the MaskEnc and MixEnc use a 4-layer 2D convolutional architecture. $H_1$ and $H_2$ are fed into the RNN-T module, and produce hypotheses $\hat{Y}_1$ and $\hat{Y}_2$ under the HEAT assumption (i.e., $Y_1$ starts before $Y_2$). The final loss is given as
\begin{equation}
\label{eq:heat}
\mathcal{L}_{\text{heat}}(X, Y_1, Y_2) = \mathcal{L}_{\text{rnnt}}(Y_1, H_1) + \mathcal{L}_{\text{rnnt}}(Y_2, H_2),
\end{equation}
where $\mathcal{L}_{\text{rnnt}}$ is the standard RNN-T loss. Although Lu et al. only evaluated HEAT on 2-speaker single-turn sessions, it is straightforward to extend HEAT to long-form sessions.

\vspace{-0.5em}
\subsection{HEAT vs. PIT}
\label{sec:heat_vs_pit}
\vspace{-0.5em}

\begin{figure}[t]
\begin{subfigure}{0.49\linewidth}
\centering
\includegraphics[width=\linewidth]{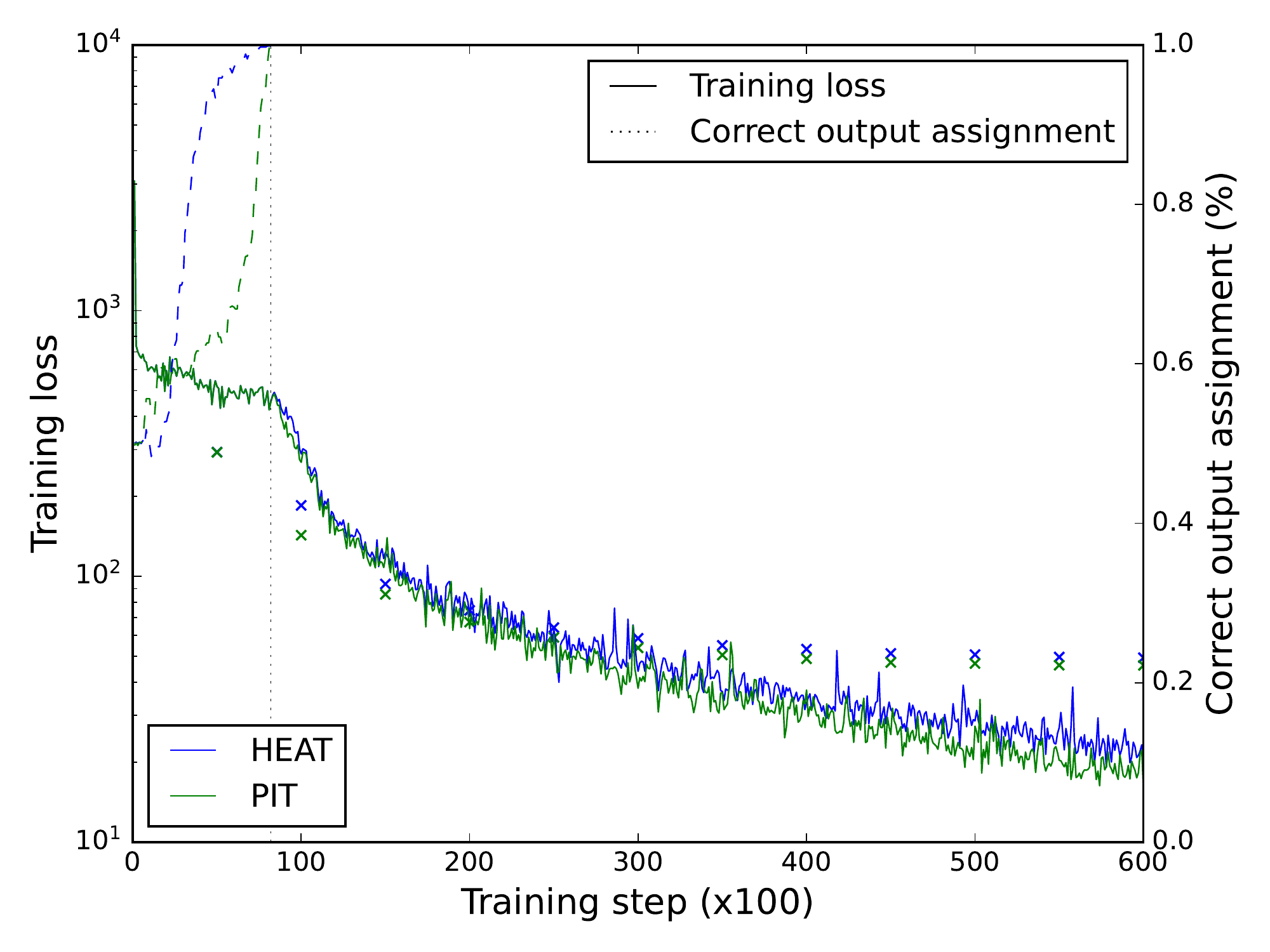}
\caption{Delay = 2.0 s}
\label{fig:delay2}
\end{subfigure}
\begin{subfigure}{0.49\linewidth}
\centering
\includegraphics[width=\linewidth]{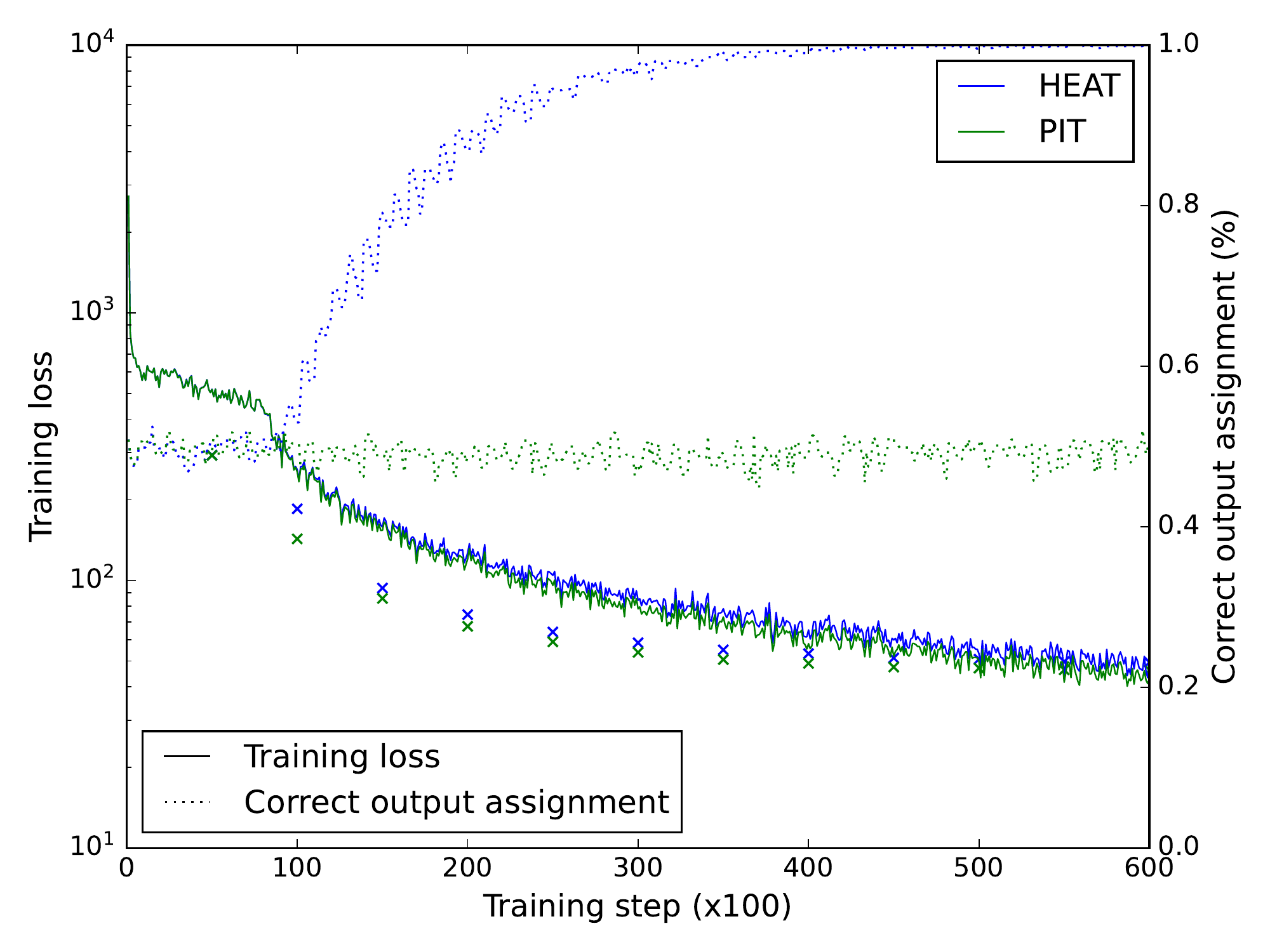}
\caption{Delay = 0.0 s}
\label{fig:delay0}
\end{subfigure}\hfill
\vspace{-0.5em}
\caption{Training dynamics for HEAT versus PIT based loss for different utterance delays: (a) 2.0 s, and (b) 0.0 s.}
\label{fig:heat_vs_pit}
\vspace{-1em}
\end{figure}

The output label permutation problem in overlapped ASR is often addressed using utterance-level permutation invariant training (PIT), which computes the optimal permutation of reference and hypothesis sequences as the training loss. For the SURT problem formulated above, it is given as
\begin{align}
\label{eq:pit}
\mathcal{L}_{\text{pit}}(X, Y_1, Y_2) = \min&\big(\mathcal{L}_{\text{rnnt}}(Y_1, H_1) + \mathcal{L}_{\text{rnnt}}(Y_2, H_2),\nonumber \\        &\mathcal{L}_{\text{rnnt}}(Y_1, H_2) + \mathcal{L}_{\text{rnnt}}(Y_2, H_1)\big).
\end{align}

In real use cases, we usually have information (such as pitch or gender) which can disambiguate the label sequences. For long-form meetings with partially overlapped utterances, utterance start time can be a good heuristic for output matching, and has been shown to outperform PIT~\cite{Lu2021StreamingEM}. To investigate this phenomenon further, we set up simple experiments on 2-utterance mixtures generated from LibriSpeech \texttt{train-clean} set. We prepared two kinds of mixtures with utterance delays of 2.0 and 0.0 seconds, respectively, and trained the vanilla SURT model using both HEAT and PIT losses (equations \ref{eq:heat} and \ref{eq:pit}, respectively). In Fig.~\ref{fig:heat_vs_pit}, we show the training dynamics for both the experiments and also plot the \% correct output assignment. This quantity represents how often the model assigns $Y_1$ to output channel 1.

As expected, for the case of mixtures with 2.0s delay, HEAT quickly learned the output assignment order. In fact, even when training with PIT, the same heuristic was learned (albeit slower), and both models started to converge only after this point was reached (denoted by the vertical line in Fig.~\ref{fig:delay2}. Thereafter, using PIT is wasteful, especially in our case of the expensive RNN-T loss computation. In the absence of utterance delay (Fig.~\ref{fig:delay0}), PIT produced a random output assignment. Surprisingly, HEAT still learned the correct assignment, but on decoding with the trained model, we found that it learned a degenerate solution where both output channels produce the exact same hypothesis.

Besides HEAT and PIT being empirically equivalent for non-zero delays, it is infeasible to compute all permutations when training on longer sessions, since the complexity grows as $\mathcal{O}(N!)$, with $N$ being the number of utterances. Even if we use linear sum assignment (which has a complexity of $\mathcal{O}(N^3)$~\cite{Kuhn1955TheHM}), it still requires computing the loss between all $N^2$ pairs of references and hypotheses, which is computationally prohibitive when using RNN-T since it implicitly sums over all possible alignments. For these reasons, we use HEAT for all experiments in the remainder of this paper. 

\begin{table}[t]
\centering
\caption{Summary of evaluation sets used in this work.}
\label{tab:eval_data}
\vspace{-0.5em}
\begin{adjustbox}{width=\linewidth} 
\begin{tabular}{@{}llcccc@{}}
\toprule
\textbf{Name}    & \textbf{Description}   & \textbf{\# spk.} & \textbf{\# utt.} & \texttt{dev} & \texttt{test} \\
\midrule
Tier-1 & 2-speaker single-turn    & 2           & 2             & 1355            & 1310             \\
Tier-2 & 2-speaker multi-turn     & 2           & 2-4           & 892             & 885              \\
Tier-3 & Multi-speaker multi-turn & 2-4         & 2-12          & 462             & 450         \\
\bottomrule
\end{tabular}
\end{adjustbox}
\vspace{-1.2em}
\end{table}

\vspace{-0.5em}
\section{Multi-turn evaluation}
\label{sec:multi-turn}
\vspace{-0.8em}


Our objective is to investigate SURT for long-form multi-talker sessions. For this, we first prepared 3 tiers of evaluation data with increasing difficulty, all of which are obtained by mixing LibriSpeech test set utterances. Tier-1 (T1) contains single-turn sessions similar to the original SURT evaluation data~\cite{Lu2021StreamingEM}. Tier-2 (T2) extends this to sessions containing up to 4 utterances, while retaining 2 speakers. Finally, Tier-3 (T3) contains multi-turn sessions with up to 4 speakers. All the tiers were generated to contain an overlap ratio between 0\% and 40\% per session. Table~\ref{tab:eval_data} summarizes the statistics for these evaluation sets. For training, we prepared 2-speaker single and multi-turn sessions which are comparable to the T1 and T2 evaluation sets, respectively. It is infeasible to train on longer sessions (like T3) because of memory constraints. Throughout this paper, we evaluated our models using the word error rate (WER) metric computed between the best permutation of reference utterances and the output channels\footnote{\label{note:wer}Since the utterances are ordered by start time, this process elicits $2^N$ permutations, which makes evaluation infeasible for $N$ greater than 12.}. For our preliminary experiments, we considered the vanilla SURT model which uses a 6-layer 1024-dim LSTM network in the RNN-T encoder (see yellow box in Fig.~\ref{fig:surt}). The prediction network is a 2-layer 1024-dim LSTM model. The model performance when trained on single and multi-turn sessions are shown below:

\begin{table}[H]
\vspace{-0.5em}
\centering
\begin{adjustbox}{width=0.65\linewidth} 
\begin{tabular}{@{}lccc@{}}
\toprule
\textbf{Train \textbackslash Eval} & \textbf{Tier-1} & \textbf{Tier-2} & \textbf{Tier-3} \\
\midrule
Single-turn & 11.1 & 17.6 & 24.9 \\
Multi-turn & 13.6 & 15.9 & 20.9 \\
\bottomrule
\end{tabular}
\end{adjustbox}
\vspace{-0.5em}
\end{table}

\textbf{Can the vanilla SURT model trained on single-turn sessions generalize to multi-turn evaluation?} They cannot. WER degrades significantly as tier complexity increases (row 1), presumably because the model had never seen any instances with multiple utterances in an output channel.

\textbf{Can we generalize to multi-turn sessions by simply training on matching data?} Although the WER on T2 and T3 improved, such training degraded T1 results (row 2), suggesting that it is challenging to make standard LSTM models generalize to diverse session lengths. An alternate model may be required which is better suited to this problem.

\vspace{-0.5em}
\section{Streaming Dual-path Transducers}
\label{sec:dual-path}
\vspace{-0.5em}

Single channel time-domain speech separation, which requires modeling extremely long sequences, has benefited from the application of dual-path (DP) networks such as DP-RNN~\cite{Luo2020DualPathRE} and DP-Transformers~\cite{Chen2020DualPathTN}. Here, we investigate streaming versions of these models as encoders in the SURT RNN-T module.

\vspace{-0.5em}
\subsection{Dual-path LSTM}
\vspace{-0.5em}

\begin{figure}[t]
\centering
\includegraphics[width=\linewidth]{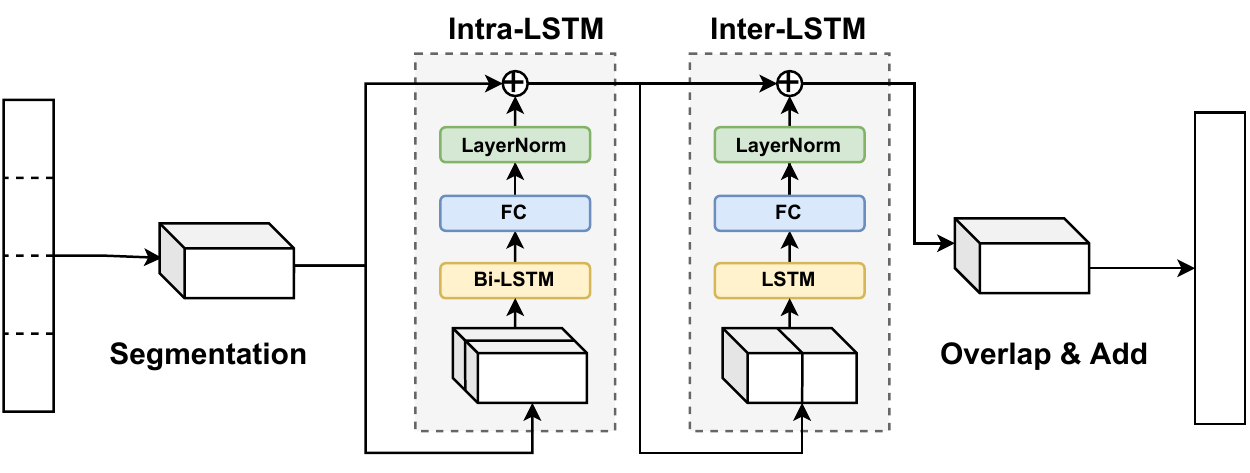}
\caption{Overview of the streaming dual path LSTM model. The intra-chunk LSTM is bidirectional, whereas the inter-chunk LSTM is unidirectional.}
\label{fig:dp_lstm_model}
\vspace{-1em}
\end{figure}

DP-LSTM consists of an \textit{intra} and an \textit{inter} LSTM per layer~\cite{Luo2020DualPathRE}, as shown in Fig.~\ref{fig:dp_lstm_model}. Input sequences are segmented into (overlapping) chunks and first fed into the bidirectional intra-LSTM, which processes each chunk independently. The output is then passed into the inter-LSTM which is unidirectional and performs strided processing over chunks. By choosing the chunk width to be approximately square root of the sequence length $l$, we can ensure that both the LSTMs get similar length inputs. Since the intra-LSTM is bidirectional, a latency equal to the chunk width is introduced in this model.

\vspace{-0.5em}
\subsection{Dual-path Transformer}
\vspace{-0.5em}

\begin{figure}[t]
\begin{subfigure}{0.24\linewidth}
\centering
\includegraphics[width=\linewidth]{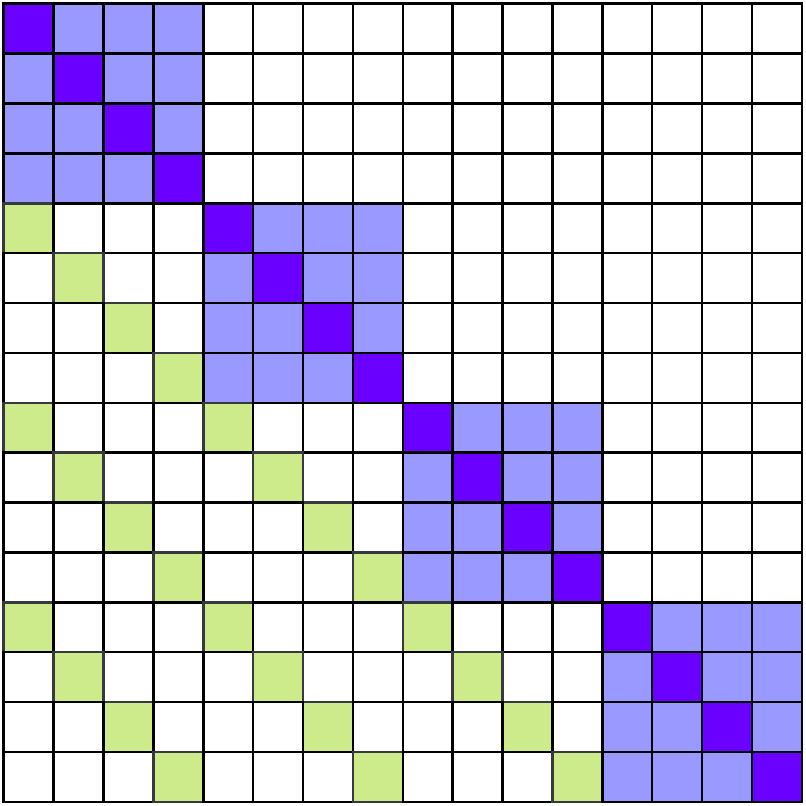}
\caption{Dual-path}
\label{fig:dp}
\end{subfigure}\hfill
\begin{subfigure}{0.24\linewidth}
\centering
\includegraphics[width=\linewidth]{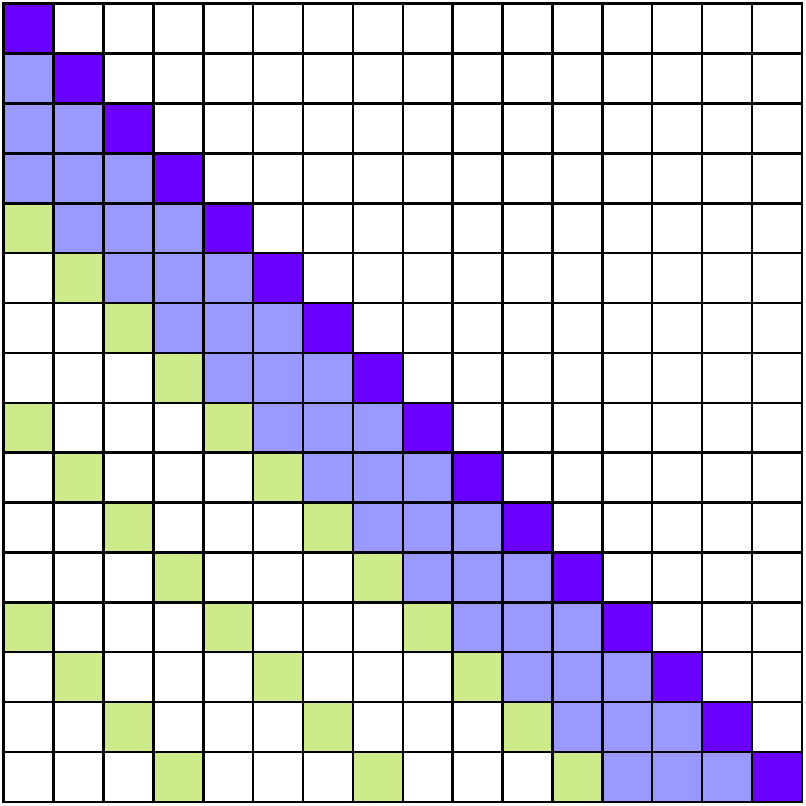}
\caption{Strided~\cite{Child2019GeneratingLS}}
\label{fig:strided}
\end{subfigure}\hfill
\begin{subfigure}{0.24\linewidth}
\centering
\includegraphics[width=\linewidth]{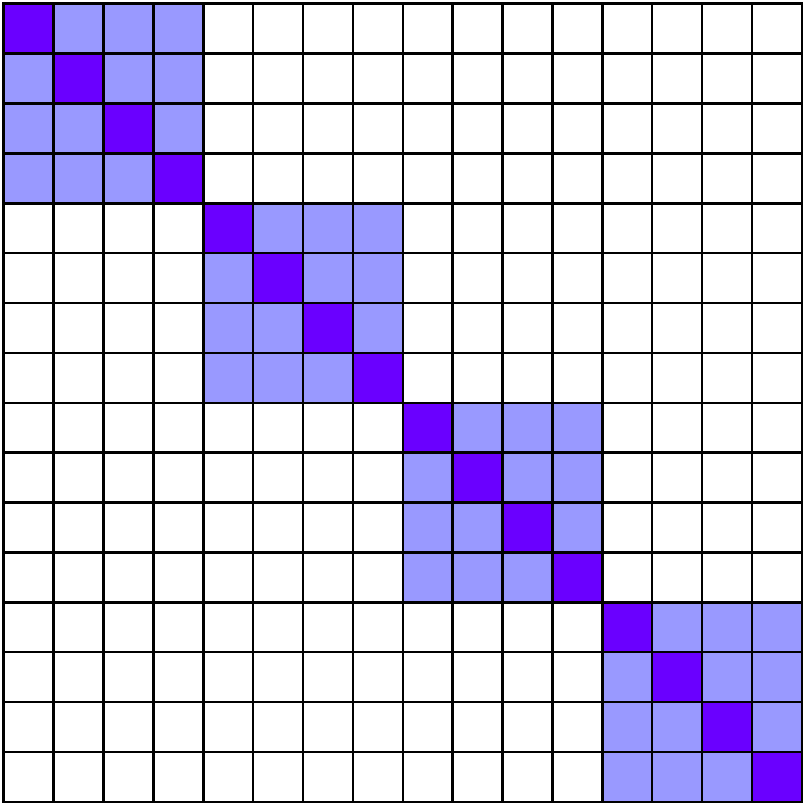}
\caption{Block~\cite{Qiu2020Blockwise}}
\label{fig:block}
\end{subfigure}\hfill
\begin{subfigure}{0.24\linewidth}
\centering
\includegraphics[width=\linewidth]{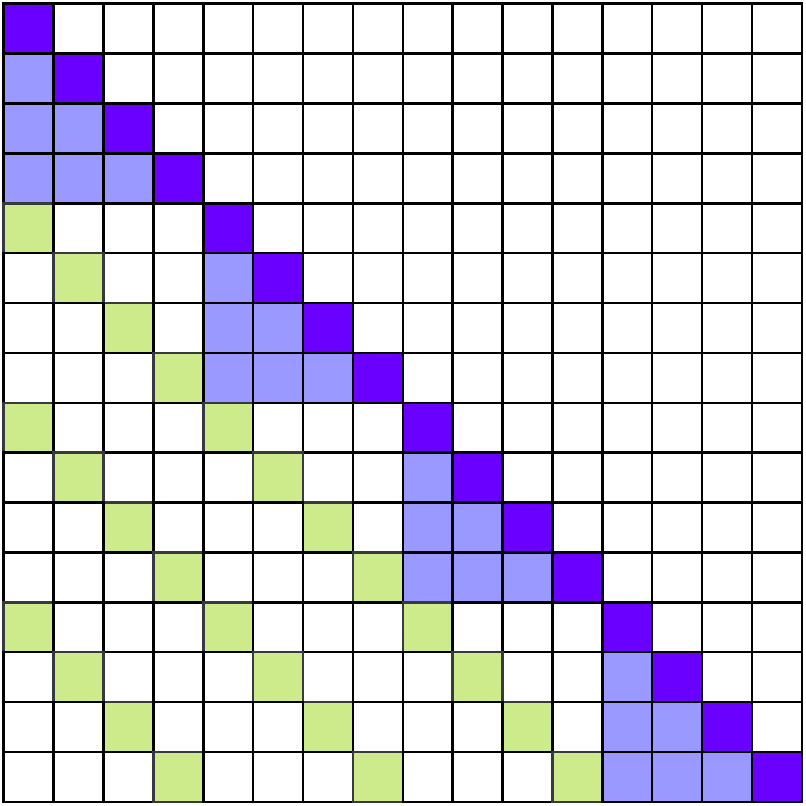}
\caption{Axial~\cite{Ho2019Axial}}
\label{fig:axial}
\end{subfigure}\hfill
\vspace{-0.5em}
\caption{Comparison of self-attention patterns in different $\mathcal{O}(l\sqrt{l})$ transformer architectures.}
\label{fig:attention}
\vspace{-0.5em}
\end{figure}

On the surface, DP-Transformer is similar to the DP-LSTM model, with the LSTM blocks replaced with self-attention blocks~\cite{Chen2020DualPathTN}. The intra-Transformer uses a full attention matrix, while the inter-Transformer is constrained to use causal attention. Consequently, the overall attention pattern for the sequence can be visualized as shown in Fig.~\ref{fig:dp}. Similar to the DP-LSTM, we can ensure $\mathcal{O}(l)$ attention computation for each chunk by choosing $\mathcal{O}(\sqrt{l})$ chunk sizes. Consequently, the DP-Transformer has a complexity of $\mathcal{O}(l\sqrt{l})$, as opposed to the quadratic complexity of the vanilla Transformer. Sparse transformers with similar attention patterns have previously been proposed for long sequences (Fig.\ref{fig:attention}). 

Despite this reduced complexity, the DP-Transformer (in non-streaming mode) is a universal function approximator. To see this, consider the following properties for the DP-Transformer. First, every token in the sequence attends to itself. Second, the directed graph $\mathcal{D}$ corresponding to the attention pattern contains a Hamiltonian path through all tokens.
Third, any token can directly/indirectly access all other tokens after exactly 2 layers (in non-streaming mode). Therefore, the model satisfies all necessary conditions for universal approximability of sparse transformers~\cite{yun2020on}.


\begin{table}[t]
\centering
\caption{WER results for SURT models with regular and dual-path encoders, trained on multi-turn data with curriculum learning.}
\label{tab:wer}
\begin{adjustbox}{width=\linewidth} 
\begin{tabular}{@{}lccccccc}
\toprule
\multirow{2}{*}{\textbf{Encoder}} & \multirow{2}{*}{\textbf{Size}} & \multicolumn{2}{c}{\textbf{Tier-1}} & \multicolumn{2}{c}{\textbf{Tier-2}} & \multicolumn{2}{c}{\textbf{Tier-3}} \\
\cmidrule(r{4pt}){3-4} \cmidrule(lr){5-6} \cmidrule(l{4pt}){7-8}
 & & \texttt{dev} & \texttt{test} & \texttt{dev} & \texttt{test} & \texttt{dev} & \texttt{test} \\
\midrule
LSTM & 75.6 M & 13.6 & 13.8 & 15.9 & 17.1 & 20.9 & 21.0 \\
DP-LSTM & \multirow{2}{*}{65.4 M} & \textbf{11.1} & 11.7 & 13.7 & 14.7 & 19.9 & 20.2 \\
~+ CWR &  & \textbf{11.1} & \textbf{11.4} & \textbf{13.0} & \textbf{14.1} & 19.6 & 19.6 \\
DP-Transformer & \multirow{2}{*}{42.9 M} & 11.5 & 12.4 & 13.7 & 15.1 & 19.1 & 20.4 \\
~+ CWR & & \textbf{11.1} & 12.2 & 13.5 & 14.5 & \textbf{17.9} & \textbf{18.6} \\
\bottomrule
\end{tabular}
\end{adjustbox}
\vspace{-1em}
\end{table}

\vspace{-0.5em}
\subsection{Chunk width randomization}
\vspace{-0.5em}

Dual-path models trained with a fixed chunk width may not be suitable for evaluation on diverse sequence lengths due to mismatch in train-test input size for the inter block. We propose training with chunk width randomization (CWR), wherein we vary the CW between a minimum and maximum value for each mini-batch. CWR increases the train time diversity in sequence length for both the intra and inter blocks and makes the model robust to such variations at test time. 

\vspace{-0.5em}
\section{Experimental setup}
\vspace{-0.8em}

The architecture of the vanilla SURT model is detailed in \S\ref{sec:multi-turn}. For the DP-based models, we replaced the RNN-T encoder with the corresponding dual path modules: a 6-layer DP-LSTM with 512-dim intra and inter blocks, and a 12-layer DP-Transformer consisting of 256-dim self-attention split into 8 heads with a 1024-dim feedforward layer. All models were trained to convergence using a learning rate schedule consisting of 10k steps of warmup to 0.0003 and a linear decay thereafter. We used AdamW as the optimizer and clipped gradients to a norm of 5. The models were trained on 16 V100 GPUs using FP16 precision. We used decoding beams of size 4 and 8 for the \texttt{dev} and \texttt{test} sets, respectively, in all experiments. For the fixed chunk dual-path models, a chunk width of 30 was used for training and decoding. For models with CWR, we trained with chunk widths between 15 and 45 and decoded with chunks of width 35. These values were tuned on the \texttt{dev} set. For the LibriCSS experiments, since the data contains far-field recordings~\cite{Chen2020ContinuousSS}, we additionally trained using simulated noise and reverberations~\cite{Kanda2021EndtoEndSA}. For these evaluations, we decoded on the approx. 1-minute long segments provided with the corpus (cf. note~\ref{note:wer}).

\vspace{-0.5em}
\section{Results \& discussion}
\vspace{-0.8em}

\noindent
\textbf{Multi-turn evaluation.}
Table~\ref{tab:wer} shows the results for our DP models compared to the LSTM-based SURT. Both the DP-LSTM and DP-Transformer provide WER improvements across all evaluation tiers without degrading performance on single-turn sessions (\S\ref{sec:multi-turn}). CWR training helped both models, with substantial gains for DP-Transformer on T3 evaluation. DP-Transformer are competitive with DP-LSTM, and outperform them on T3, with only two-thirds the model size.


\begin{figure}[t]
    \centering
    \includegraphics[width=0.55\linewidth]{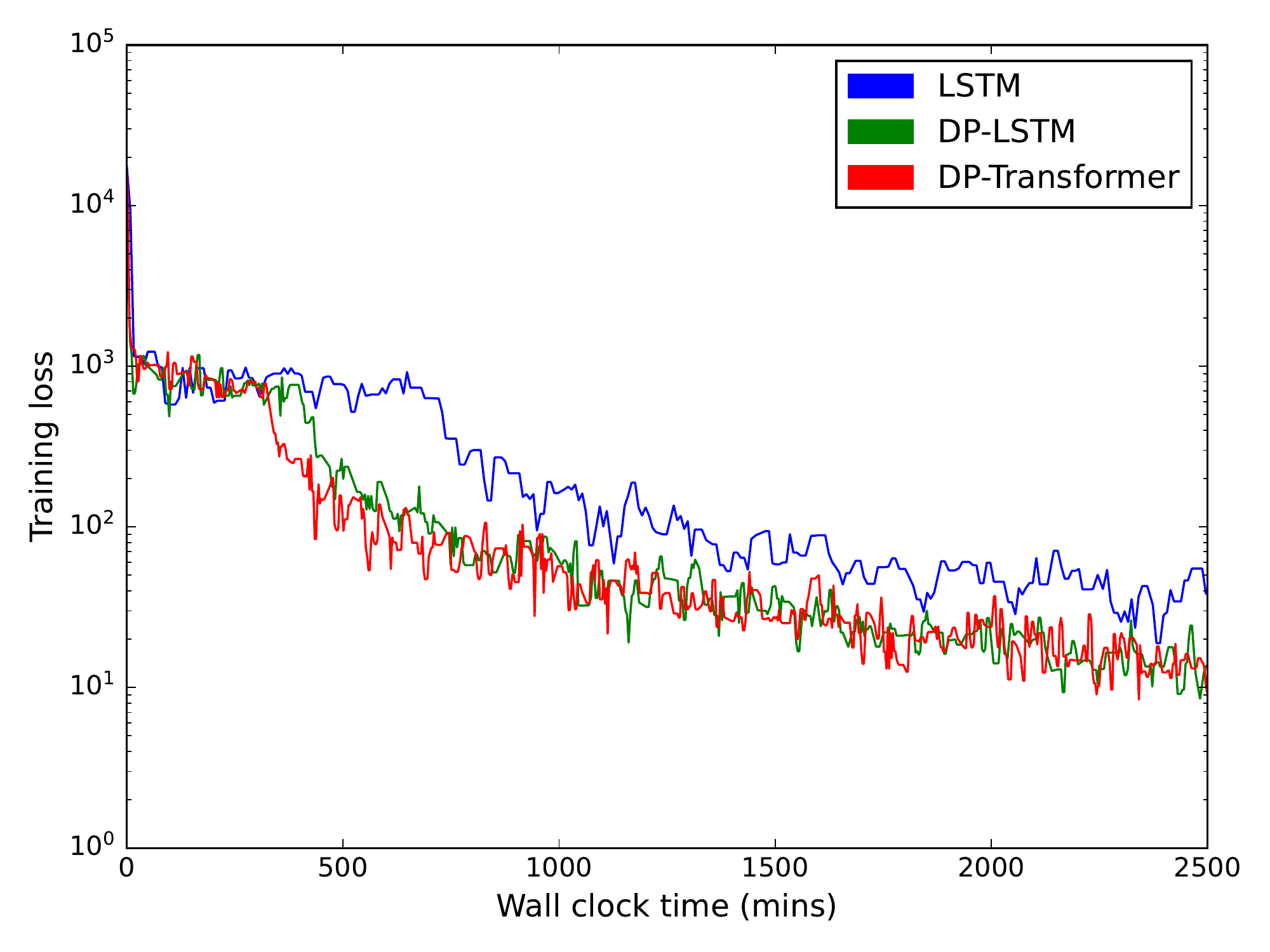}
    \vspace{-0.5em}
    \caption{Training curves for the dual-path models compared with the regular LSTM model, plotted w.r.t. wall clock time (mins).}
    \label{fig:training_curve}
    \vspace{-1.5em}
\end{figure}

\begin{table}[b]
\centering
\vspace{-1.5em}
\caption{Effect of curriculum learning on \texttt{dev} set. We compare models trained only on multi-turn sessions versus those initialized from 20k steps of single-turn training.}
\label{tab:curriculum}
\vspace{-0.5em}
\begin{adjustbox}{width=\linewidth} 
\begin{tabular}{@{}lcccccc}
\toprule
\multirow{2}{*}{\textbf{Model}} & \multicolumn{2}{c}{\textbf{Tier-1}} & \multicolumn{2}{c}{\textbf{Tier-2}} & \multicolumn{2}{c}{\textbf{Tier-3}} \\
\cmidrule(r{4pt}){2-3} \cmidrule(lr){4-5} \cmidrule(l{4pt}){6-7}
 & \textit{None} & \textit{20k} & \textit{None} & \textit{20k} & \multicolumn{1}{c}{\textit{None}} & \multicolumn{1}{c}{\textit{20k}} \\
\midrule
LSTM & 15.8 & 13.6 {\footnotesize($\downarrow$14.1\%)} & 17.8 & 15.9 {\footnotesize($\downarrow$10.7\%)} & 21.8 & 20.9 {\footnotesize($\downarrow$4.1\%)} \\
DP-LSTM & 11.4 & 11.1 {\footnotesize($\downarrow$2.6\%)} & 13.6 & 13.0 {\footnotesize($\downarrow$4.4\%)} & 18.6 & 19.6 {\footnotesize($\uparrow$5.4\%)} \\
DP-Transformer & 12.5 & 11.1 {\footnotesize($\downarrow$11.2\%)} & 13.7 & 13.5 {\footnotesize($\downarrow$1.5\%)} & 18.2 & 17.9 {\footnotesize($\downarrow$1.6\%)} \\
\bottomrule
\end{tabular}
\end{adjustbox}
\end{table}

\noindent
\textbf{Training dynamics and curriculum learning.}
Fig.~\ref{fig:training_curve} shows the training curves for the dual-path models compared with the LSTM-based SURT model. Both the DP-LSTM and DP-Transformer converged faster and to a better minimum, despite smaller model sizes. We also found it important to use a curriculum learning strategy for all the models, where we train them first on single-turn sessions, and then introduce multi-turn sessions after 20k training steps. Table~\ref{tab:curriculum} compares performances on the \texttt{dev} set for models trained with and without a curriculum. Since the multi-turn training data also contains single-turn sessions, and the benefits are seen across all tiers, these improvements cannot be attributed solely to variation in training data.

\begin{figure}[t]
\begin{subfigure}{0.49\linewidth}
\centering
\includegraphics[width=\linewidth]{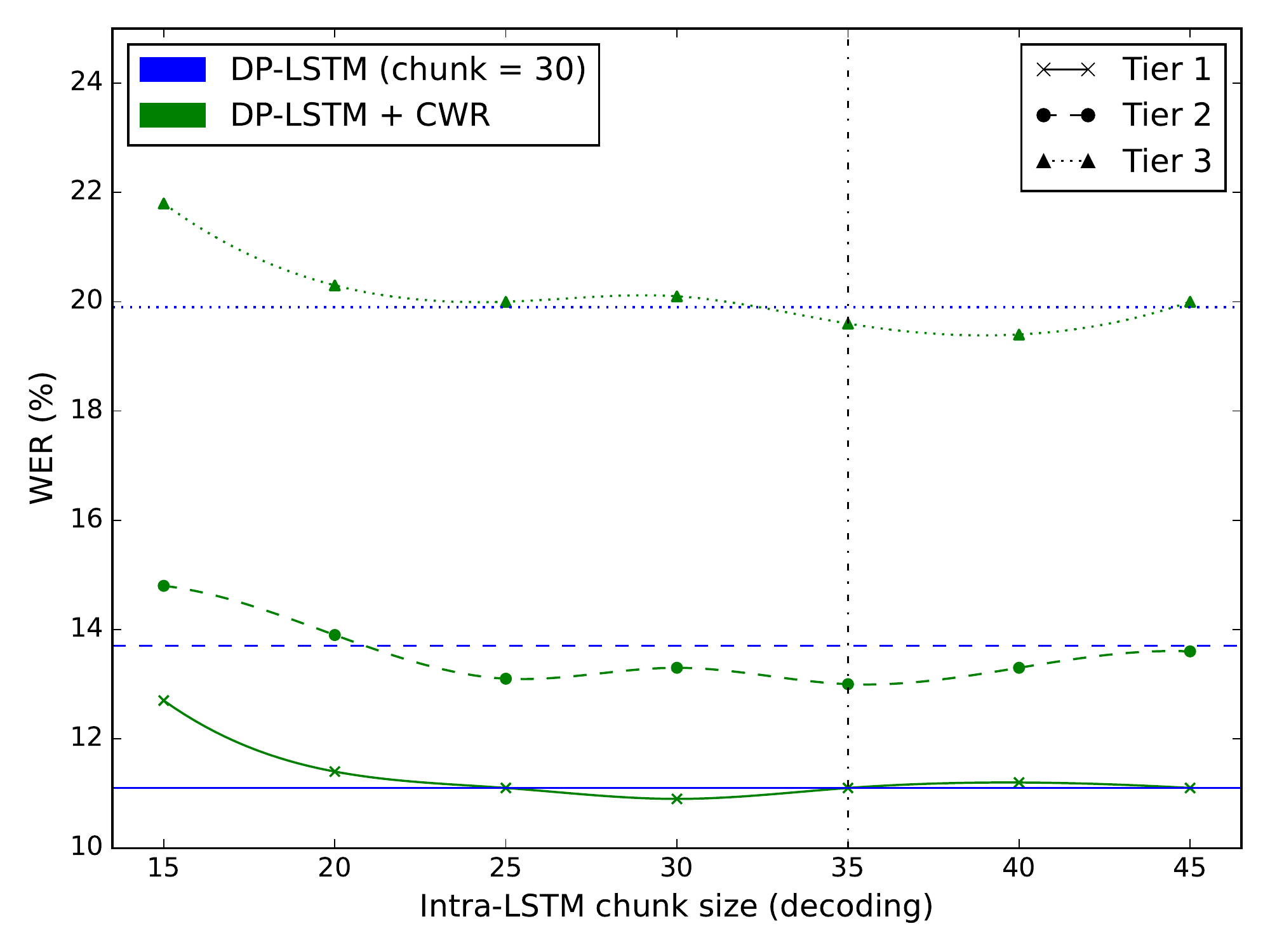}
\caption{DP-LSTM}
\label{fig:dp_lstm_chunk}
\end{subfigure}
\begin{subfigure}{0.49\linewidth}
\centering
\includegraphics[width=\linewidth]{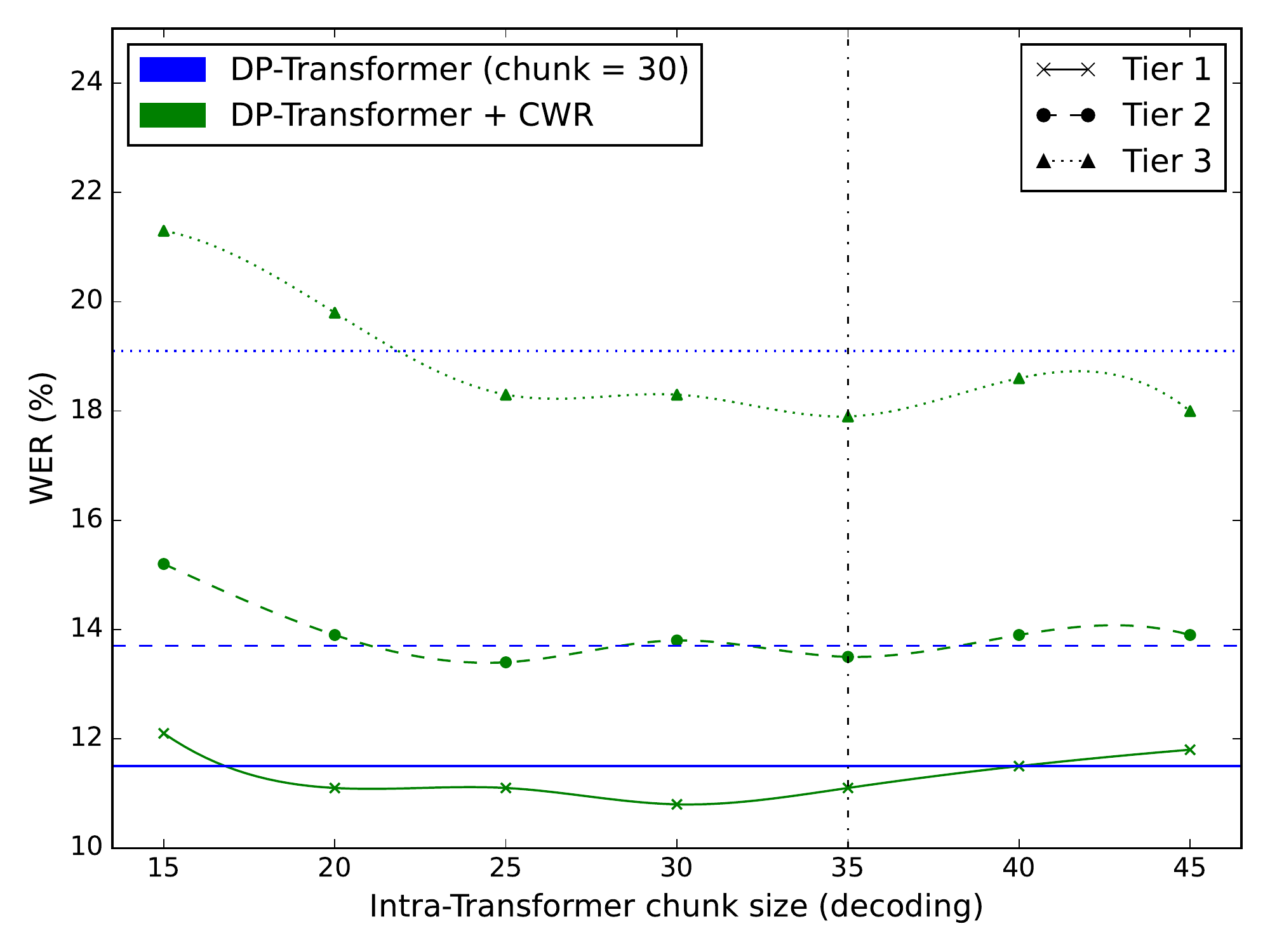}
\caption{DP-Transformer}
\label{fig:dp_transformer_chunk}
\end{subfigure}\hfill
\vspace{-0.5em}
\caption{Accuracy vs. latency trade-off for dual-path models trained with chunk width randomization (CWR), evaluated on \texttt{dev} set.}
\label{fig:chunk}
\end{figure}

\noindent
\textbf{Accuracy vs. latency.}
Besides improving robustness to session lengths, CWR also enables models to be deployed with lower latency at the cost of a small WER degradation. Fig.~\ref{fig:chunk} shows this trade-off between decoding latency (determined by chunk width) and WER performance. WER tends to degrade slightly for low latency decoding, and the best performance is obtained for chunk sizes of 35.

\noindent
\textbf{Evaluation on LibriCSS.}
Table~\ref{tab:libricss} compares our models against {\it offline} modular systems composed of continuous speech separation (CSS) and ASR modules. The ``conformer CSS + E2E ASR'' model~\cite{Chen2021ContinuousSS} contains 197M parameters (59M for CSS, and 138M for ASR~\cite{Wang2020SemanticMF}), and uses language model fusion and rescoring. On high overlap conditions, our models are competitive with these state-of-the-art offline modular systems, with most errors arising from \textit{leakage} in single-speaker regions, i.e., when both channels transcribed an utterance even when no overlap occurred. For the low overlap cases, \textit{omissions} were another major error source, where some utterances were completely missed by both channels. DP-Transformer does not outperform DP-LSTM in this case possibly due to overfitting to the simulated noise and reverberation, as evident from their performance in the \textit{clean} (digitally mixed clean audio) setting. 

\begin{table}[t]
\centering
\caption{WER results for single-channel LibriCSS. 0L and 0S denote 0\% overlap with long and short silences, respectively.}
\label{tab:libricss}
\vspace{-0.5em}
\begin{adjustbox}{max width=\linewidth}
\begin{tabular}{@{}lccccccc@{}}
\toprule
\multicolumn{2}{l}{\multirow{2}{*}{\textbf{Model}}} & \multicolumn{6}{c}{\textbf{Overlap ratio in \%}} \\
\cmidrule(r{4pt}){3-8}
\multicolumn{2}{c}{} & \multicolumn{1}{c}{\textbf{0L}} & \multicolumn{1}{c}{\textbf{0S}} & \multicolumn{1}{c}{\textbf{10}} & \multicolumn{1}{c}{\textbf{20}} & \multicolumn{1}{c}{\textbf{30}} & \multicolumn{1}{c}{\textbf{40}} \\
\midrule
\multicolumn{2}{l}{BLSTM CSS + Hybrid ASR~\cite{Chen2020ContinuousSS}} & 16.3 & 17.6 & 20.9 & 26.1 & 32.6 & 36.1 \\
\multicolumn{2}{l}{Conformer CSS + E2E ASR~\cite{Chen2021ContinuousSS}} & 6.1 & 6.9 & 9.1 & 12.5 & 16.7 & 19.3 \\
\midrule
SURT w/ DP-LSTM & \textit{replayed} & 9.8 & 19.1 & 20.6 & 20.4 & 23.9 & 26.8  \\
\rowcolor{Gray}
SURT w/ DP-LSTM & \textit{clean} & 6.6 & 21.6 & 21.7 & 20.6 & 25.4 & 28.4 \\
SURT w/ DP-Transformer & \textit{replayed} & 9.3 & 21.1 & 21.2 & 25.9 & 28.2 & 31.7 \\
\rowcolor{Gray}
SURT w/ DP-Transformer & \textit{clean} & 6.9 & 18.9 & 19.6 & 21.9 & 23.9 & 28.7 \\
\bottomrule
\end{tabular}
\end{adjustbox}
\vspace{-1em}
\end{table}


\vspace{-0.5em}
\section{Conclusion}
\vspace{-0.8em}

We investigated SURT for continuous streaming multi-talker ASR, and demonstrated the effectiveness of dual-path LSTMs and Transformers for generalization to diverse session lengths. Training strategies such as curriculum learning and chunk width randomization provided WER improvements and enabled the model to be deployed with different latencies. We also showed that DP-Transformers are universal function approximators. Our models demonstrated encouraging results across different overlap levels on the LibriCSS dataset while being smaller, faster, and simpler than modular systems. In future work, we will explore segmentation in the SURT output channels, which will naturally enable speaker-attributed ASR.

\small
\bibliographystyle{IEEEbib}
\bibliography{refs}

\begin{thebibliography}{10}

\bibitem{Amodei2016DS2}
D.~Amodei et~al.,
\newblock ``Deep speech 2 : End-to-end speech recognition in english and
  mandarin,''
\newblock in {\em Proceedings of The 33rd International Conference on Machine
  Learning}, 2016.

\bibitem{Xiong2017TowardHP}
W.~Xiong, J.~Droppo, X.~Huang, F.~Seide, M.~Seltzer, A.~Stolcke, D.~Yu, and
  G.~Zweig,
\newblock ``Toward human parity in conversational speech recognition,''
\newblock {\em IEEE/ACM Transactions on Audio, Speech, and Language
  Processing}, 2017.

\bibitem{Barker2015TheT}
J.~Barker, R.~Marxer, E.~Vincent, and S.~Watanabe,
\newblock ``The third {‘CHiME’} speech separation and recognition
  challenge: Dataset, task and baselines,''
\newblock in {\em IEEE ASRU}, 2015.

\bibitem{Kinoshita2013TheRC}
K.~Kinoshita, M.~Delcroix, T.~Yoshioka, T.~Nakatani, A.~Sehr, W.~Kellermann,
  and R.~Maas,
\newblock ``The {REVERB} challenge: A common evaluation framework for
  dereverberation and recognition of reverberant speech,''
\newblock in {\em IEEE WASPAA}, 2013.

\bibitem{Watanabe2020CHiME6CT}
S.~Watanabe, M.~Mandel, J.~Barker, and E.~Vincent,
\newblock ``{CHiME-6} challenge: Tackling multispeaker speech recognition for
  unsegmented recordings,''
\newblock {\em ArXiv}, vol. abs/2004.09249, 2020.

\bibitem{Carletta2005TheAM}
J.~Carletta et~al.,
\newblock ``The {AMI} meeting corpus: A pre-announcement,''
\newblock in {\em MLMI}, 2005.

\bibitem{Shriberg2001ObservationsOO}
E.~Shriberg, A.~Stolcke, and D.~Baron,
\newblock ``Observations on overlap: findings and implications for automatic
  processing of multi-party conversation,''
\newblock in {\em INTERSPEECH}, 2001.

\bibitem{Yoshioka2019MeetingTU}
T.~Yoshioka, D.~Dimitriadis, A.~Stolcke, W.~Hinthorn, Z.~Chen, M.~Zeng, and
  X.~Huang,
\newblock ``Meeting transcription using asynchronous distant microphones,''
\newblock in {\em INTERSPEECH}, 2019.

\bibitem{Graves2006ConnectionistTC}
A.~Graves, S.~Fern{\'a}ndez, F.~Gomez, and J.~Schmidhuber,
\newblock ``Connectionist temporal classification: labelling unsegmented
  sequence data with recurrent neural networks,''
\newblock in {\em ICML}, 2006.

\bibitem{Graves2012SequenceTW}
A.~Graves,
\newblock ``Sequence transduction with recurrent neural networks,''
\newblock in {\em ICML Representation Learning Workshop}, 2012.

\bibitem{Lu2016OnTT}
L.~Lu, X.~Zhang, and S.~Renals,
\newblock ``On training the recurrent neural network encoder-decoder for large
  vocabulary end-to-end speech recognition,''
\newblock in {\em IEEE ICASSP}, 2016.

\bibitem{Chorowski2015AttentionBasedMF}
J.~Chorowski, D.~Bahdanau, D.~Serdyuk, K.~Cho, and Y.~Bengio,
\newblock ``Attention-based models for speech recognition,''
\newblock in {\em Advances in Neural Information Processing Systems}, 2015.

\bibitem{Chiu2018StateoftheArtSR}
C.~Chiu, T.~Sainath, Y.~Wu, R.~Prabhavalkar, P.~Nguyen, Z.~Chen, A.~Kannan,
  R.~J. Weiss, K.~Rao, K.~Gonina, N.~Jaitly, B.~Li, J.~Chorowski, and
  M.~Bacchiani,
\newblock ``State-of-the-art speech recognition with sequence-to-sequence
  models,''
\newblock in {\em IEEE ICASSP}, 2018.

\bibitem{he2019streaming}
Y.~He, T.~N. Sainath, R.~Prabhavalkar, I.~McGraw, R.~Alvarez, D.~Zhao,
  D.~Rybach, A.~Kannan, Y.~Wu, R.~Pang, et~al.,
\newblock ``Streaming end-to-end speech recognition for mobile devices,''
\newblock in {\em IEEE ICASSP}, 2019.

\bibitem{Li2019RNNT}
J.~Li, R.~Zhao, H.~Hu, and Y.~Gong,
\newblock ``Improving {RNN} transducer modeling for end-to-end speech
  recognition,''
\newblock in {\em IEEE ASRU}, 2019.

\bibitem{li2021recent}
J.~Li,
\newblock ``Recent advances in end-to-end automatic speech recognition,''
\newblock {\em arXiv preprint arXiv:2111.01690}, 2021.

\bibitem{Settle2018EndtoEndMS}
S.~Settle, J.~Le Roux, T.~Hori, S.~Watanabe, and J.~Hershey,
\newblock ``End-to-end multi-speaker speech recognition,''
\newblock in {\em IEEE ICASSP}, 2018.

\bibitem{Chang2019EndtoendMM}
X.~Chang, Y.~Qian, K.~Yu, and S.~Watanabe,
\newblock ``End-to-end monaural multi-speaker {ASR} system without
  pretraining,''
\newblock in {\em IEEE ICASSP}, 2019.

\bibitem{Kanda2020SerializedOT}
N.~Kanda, Y.~Gaur, X.~Wang, Z.~Meng, and T.~Yoshioka,
\newblock ``Serialized output training for end-to-end overlapped speech
  recognition,''
\newblock in {\em INTERSPEECH}, 2020.

\bibitem{Tripathi2020EndToEndMO}
A.~Tripathi, H.~Lu, and H.~Sak,
\newblock ``End-to-end multi-talker overlapping speech recognition,''
\newblock in {\em IEEE ICASSP}, 2020.

\bibitem{Sklyar2021StreamingMA}
I.~Sklyar, A.~Piunova, and Y.~Liu,
\newblock ``Streaming multi-speaker {ASR} with {RNN-T},''
\newblock in {\em IEEE ICASSP}, 2021.

\bibitem{Lu2021StreamingEM}
L.~Lu, N.~Kanda, J.~Li, and Y.~Gong,
\newblock ``Streaming end-to-end multi-talker speech recognition,''
\newblock {\em IEEE Signal Processing Letters}, 2021.

\bibitem{Luo2020DualPathRE}
Y.~Luo, Z.~Chen, and T.~Yoshioka,
\newblock ``Dual-path {RNN}: Efficient long sequence modeling for time-domain
  single-channel speech separation,''
\newblock in {\em IEEE ICASSP}, 2020.

\bibitem{Chen2020DualPathTN}
J-J. Chen, Q.~Mao, and D.~Liu,
\newblock ``Dual-path transformer network: Direct context-aware modeling for
  end-to-end monaural speech separation,''
\newblock in {\em INTERSPEECH}, 2020.

\bibitem{Chen2020ContinuousSS}
Z.~Chen, T.~Yoshioka, L.~Lu, T.~Zhou, Z.~Meng, Y.~Luo, J.~Wu, and J.~Li,
\newblock ``Continuous speech separation: Dataset and analysis,''
\newblock in {\em IEEE ICASSP}, 2020.

\bibitem{Kuhn1955TheHM}
H.~Kuhn,
\newblock ``The hungarian method for the assignment problem,''
\newblock {\em Naval Research Logistics Quarterly}, vol. 2, pp. 83--97, 1955.

\bibitem{Child2019GeneratingLS}
R.~Child, S.~Gray, A.~Radford, and I.~Sutskever,
\newblock ``Generating long sequences with sparse transformers,''
\newblock {\em ArXiv}, vol. abs/1904.10509, 2019.

\bibitem{Qiu2020Blockwise}
J.~Qiu, H.~Ma, O.~Levy, W-T. Yih, S.~Wang, and J.~Tang,
\newblock ``Blockwise self-attention for long document understanding,''
\newblock in {\em Findings of the Association for Computational Linguistics:
  EMNLP 2020}, 2020.

\bibitem{Ho2019Axial}
J.~Ho, N.~Kalchbrenner, D.~Weissenborn, and T.~Salimans,
\newblock ``Axial attention in multidimensional transformers,''
\newblock {\em ArXiv}, 2019.

\bibitem{yun2020on}
C.~Yun, Y-W. Chang, S.~Bhojanapalli, A.~S. Rawat, S.~Reddi, and S.~Kumar,
\newblock ``O(n) connections are expressive enough: Universal approximability
  of sparse transformers,''
\newblock in {\em Advances in Neural Information Processing Systems}, 2020.

\bibitem{Kanda2021EndtoEndSA}
N.~Kanda, G.~Ye, Y.~Gaur, X.~Wang, Z.~Meng, Z.~Chen, and T.~Yoshioka,
\newblock ``End-to-end speaker-attributed {ASR} with transformer,''
\newblock in {\em INTERSPEECH}, 2021.

\bibitem{Chen2021ContinuousSS}
S.~Chen, Y.~Wu, Z.~Chen, J.~Li, C.~Wang, S.~Liu, and M.~Zhou,
\newblock ``Continuous speech separation with conformer,''
\newblock in {\em IEEE ICASSP}, 2021.

\bibitem{Wang2020SemanticMF}
C.~Wang, Y.~Wu, Y.~Du, J.~Li, S.~Liu, L.~Lu, S.~Ren, G.~Ye, S.~Zhao, and
  M.~Zhou,
\newblock ``Semantic mask for transformer based end-to-end speech
  recognition,''
\newblock in {\em INTERSPEECH}, 2020.

\end{thebibliography}

\end{document}